\documentclass[5p]{elsarticle}

\usepackage{graphicx}
\usepackage{multirow}
\usepackage{stfloats}
\usepackage{subcaption}
\usepackage{makecell}
\usepackage{soul}
\usepackage{color, xcolor}
\soulregister{\cite}7
\soulregister\autoref7
\usepackage{amsfonts,amssymb}
\usepackage{amsmath}
\usepackage{array}
\usepackage{siunitx}
\usepackage[ruled,vlined]{algorithm2e}
\usepackage{booktabs}

\usepackage{lineno,hyperref}
\modulolinenumbers[5]

\journal{Journal of Computational Physics}

\begin{document}

\begin{frontmatter}

\title{Octree-based hierarchical sampling optimization for the volumetric super-resolution of scientific data}
% \tnotetext[mytitlenote]{Fully documented templates are available in the elsarticle package on \href{http://www.ctan.org/tex-archive/macros/latex/contrib/elsarticle}{CTAN}.}

%% Group authors per affiliation:
% \author{Elsevier\fnref{myfootnote}}
% \address{Radarweg 29, Amsterdam}
% \fntext[myfootnote]{Since 1880.}

%% or include affiliations in footnotes:
\author[a]{Xinjie Wang}
\author[a]{Maoquan Sun}
\author[b]{Yundong Guo\corref{mycorrespondingauthor}}
\cortext[mycorrespondingauthor]{Corresponding author}
\ead{guoyd@sdust.edu.cn}

\author[c]{Chunxin Yuan}
\author[c]{Xiang Sun}
\author[a]{Zhiqiang Wei}
\author[d]{Xiaogang Jin}

\address[a]{College of Computer Science and Technology, Ocean University of China, Qingdao, 266100, China}
\address[b]{College of Computer Science and Engineering, Shandong University of Science and Technology, Qingdao, 266590, China}
\address[c]{School of Mathematical Sciences, Ocean University of China, Qingdao, 266100, China}
\address[d]{State Key Lab of CAD\&CG, Zhejiang University, Hangzhou, 310058, China}

\begin{abstract}
When introducing physics-constrained deep learning solutions to the volumetric super-resolution of scientific data, the training is challenging to converge and always time-consuming. We propose a new hierarchical sampling method based on octree to solve these difficulties. In our approach, scientific data is preprocessed before training, and a hierarchical octree-based data structure is built to guide sampling on the latent context grid. Each leaf node in the octree corresponds to an indivisible subblock of the volumetric data. The dimensions of the subblocks are different, making the number of sample points in each randomly cropped training data block to be adaptive. We reconstruct the octree at intervals according to loss distribution to perform the multi-stage training. With the Rayleigh-B{\'e}nard convection problem, we deploy our method to state-of-the-art models. We constructed adequate experiments to evaluate the training performance and model accuracy of our method. Experiments indicate that our sampling optimization improves the convergence performance of physics-constrained deep learning super-resolution solutions. Furthermore, the sample points and training time are significantly reduced with no drop in model accuracy. We also test our method in training tasks of other deep neural networks, and the results show our sampling optimization has extensive effectiveness and applicability. The code is publicly available at \href{https://github.com/xinjiewang/octree-based_sampling}{https://github.com/xinjiewang/octree-based\_sampling}.
\end{abstract}

\begin{keyword}
% \texttt{elsarticle.cls}\sep \LaTeX\sep Elsevier \sep template
% \MSC[2010] 00-01\sep  99-00
scientific data\sep volumetric super-resolution\sep octree\sep importance sampling\sep physics-constrained deep learning
\end{keyword}

\end{frontmatter}

% \linenumbers

\section{Introduction}\label{sec:introduction}

% \paragraph{Installation} If the document class \emph{elsarticle} is not available on your computer, you can download and install the system package \emph{texlive-publishers} (Linux) or install the \LaTeX\ package \emph{elsarticle} using the package manager of your \TeX\ installation, which is typically \TeX\ Live or Mik\TeX.

% \paragraph{Usage} Once the package is properly installed, you can use the document class \emph{elsarticle} to create a manuscript. Please make sure that your manuscript follows the guidelines in the Guide for Authors of the relevant journal. It is not necessary to typeset your manuscript in exactly the same way as an article, unless you are submitting to a camera-ready copy (CRC) journal.

% \paragraph{Functionality} The Elsevier article class is based on the standard article class and supports almost all of the functionality of that class. In addition, it features commands and options to format the
% \begin{itemize}
% \item document style
% \item baselineskip
% \item front matter
% \item keywords and MSC codes
% \item theorems, definitions and proofs
% \item lables of enumerations
% \item citation style and labeling.
% \end{itemize}

To study complex physical phenomena (e.g., turbulence flows\cite{chung2002, Yeung2015}, ocean currents \cite{Hurrell2013, Vallis2017}, climate systems\cite{Richardson2007Weather, Almgren2013, Bauer2015weather}), scientists have to run numerical simulations on supercomputers without interruption to continuously generate high-resolution scientific data. These produced volumetric data with spatio-temporal dimensions are vital for scientists to discover and verify physical laws. It is worth noting that as the resolution of numerical simulations increases, the scale of revealed physical processes decreases. Nevertheless, this entails a greater demand for computational resources, and the simulation time becomes intolerable. It severely constrains the study of small-scale physical phenomena such as oceanic submesoscale processes.

To efficiently acquire high-resolution scientific data within a limited time, it is recommended to develop a physics-constrained deep neural network that can conduct super-resolution reconstruction \cite{Jiang2020, Arora2022}. Since it can simultaneously learn statistical and physical correlations between pairs of low-resolution and high-resolution solutions, this deep learning technique has been successfully applied in the volumetric super-resolution of flow simulations \cite{Wang2022, Ren2022}. However, during the training process, deep learning models encounter difficulty in converging within a reasonable time due to the complexity of partial differential equations (PDEs) involved. Therefore, how to enhance the training performance of physics-constrained super-resolution models still remains a challenge.

When training a super-resolution model incorporating physical constraints, dense sampling on the latent context grid is a prerequisite. The spatio-temporal coordinate of each sample point along with its implicit feature is fed into the decoder to produce the corresponding value of physical variables at the respective coordinate. As a result, the decoder performs much more times than the encoder, and the computational burden of training is greatly impacted by the number of sample points. In physics-informed neural networks (PINNs), implementing importance sampling, rather than random sampling, proves to be a viable approach \cite{Yang2022}. To this end, in this paper, we focus on optimizing the distribution of the sample points to improve the training performance of physics-constrained super-resolution models.

Some previous works have demonstrated that importance sampling is an efficient method for accelerating the training of deep neural networks \cite{katharopoulos2017biased, chen2018fastgcn, Johnson2018, banerjee2021deterministic} and PINNs \cite{nabian2021efficient, lu2021deepxde, Yang2022}. However, these methods focus on evaluating the importance weight of each training sample or sample point, ignoring the dynamic overall importance distribution of the whole data. Besides, there is currently no importance sampling procedure specifically designed and optimized for the volumetric data.

Inspired by the high efficiency of octree \cite{que2021voxelcontext, wang2022dual, fu2022octattention}, we propose an octree-based hierarchical sampling optimization method that is elaborately designed for the training of physics-constrained super-resolution models. By building an octree model of the training data, we are able to guide the sampling process based on its importance distribution. This approach leads to a notable reduction in the number of sample points and accelerates the convergence of physics-constrained deep learning models.

Specifically, our main technical contributions in this paper are as follows:
\begin{itemize}
\item We propose a novel octree-based approach for modeling volumetric scientific data by leveraging its importance distribution, wherein each leaf node corresponds to an indivisible subblock in the volumetric domain.

\item We present a hierarchical sampling optimization method, which performs importance sampling based on the octree model to improve the distribution of sample points in each training data block.

\item We construct a multi-stage training strategy to avoid failure of the proposed sampling optimization, which involves updating the octree model whenever the training encounters a bottleneck.
\end{itemize}

We apply the proposed octree-based hierarchical sampling optimization in two physics-constrained super-resolution solutions and perform extensive experiments on Rayleigh-B{\'e}nard convection problem. The results show that our approach can significantly improve the training performance of deep learning super-resolution models with physical constraints. Specifically, the octree-based sampling optimization method can help the model achieve the same accuracy with fewer sample points and a shorter time.

The rest parts of this paper are organized as follows. \autoref{sec:relatedwork} briefly reviews the related work, including volumetric super-resolution of scientific data and importance sampling methods for deep learning. \autoref{sec:method} describes the proposed method, covering five technical details comprehensively. \autoref{sec:experiments} demonstrates the detailed experimental results and \autoref{sec:limitations} gives some limitations and discussions. Finally, conclusions are drawn, and the associated future work is given in \autoref{sec:conclusion}.

\section{Related Work}\label{sec:relatedwork}
We focus on developing a sampling optimization method for the deep learning-based volumetric super-resolution. Therefore, related work includes two aspects: the volumetric super-resolution of scientific data and the importance sampling for deep learning.

\subsection{Volumetric Super-resolution of Scientific Data}
The volumetric super-resolution of scientific data aims to reconstruct the fine-scale subgrid solutions from the coarse-scale ones. It is usually implemented based on interpolation algorithms such as trilinear interpolation. In recent years, due to its impressive ability to tackle multi-dimensional and nonlinear issues, the deep learning technique has been introduced to volumetric super-resolution tasks and is widely used.

Xie et al. \cite{xie2018tempogan} proposed a conditional generative adversarial network (GAN) for the volumetric data of fluid flows. It is able to generate consistent and detailed super-resolution results by using a new temporal discriminator. Werhahn et al. \cite{Werhahn2019} obtained full coverage for the volumetric domain by two separate GANs and can leverage spatio-temporal supervision with a set of discriminators. in addition to GAN, Fukami et al. \cite{fukami2019super} employed other deep learning techniques such as convolutional neural networks and multi-scale models with skip-connections to build and super-resolve 3D flow fields. Besides, in the field of 3D visualization, SSR-VFD \cite{Guo2020} was presented to produce coherent spatial super-resolution of 3D vector field data, and TSR-TVD \cite{Han2020} was presented to generate temporal super-resolution of time-varying data using the recurrent generative network. Weiss et al. \cite{Weiss2021} also proposed a fully convolutional neural network to learn an upscaling model for rendering accurate isosurfaces in a volumetric field.

These methods can offer statistical features, but they do not consider the physical laws that scientific data must adhere to. The lack of interpretability is a common problem of deep learning, which leads to unreliable super-resolution results of scientific data. As a solution to this problem, researchers introduced PINNs \cite{Raissi2019PINN, zhu2019physics, Pang2019fPINNs, Jagtap2020XPINNs, yang2021b, jin2021nsfnets, Shukla2021Parallel, XU2022}. PINNs are neural networks that learn distribution patterns of training data to approximate the physical laws. It has been applied in a range of fields, such as materials \cite{shukla2020physics, goswami2022physics}, mechanics\cite{Haghighat2021, sharma2021physics}, fluids\cite{Wang2020Towards, mao2020physics, chen2022flowdnn} and bioengineering \cite{Sahli2020, Kissas2020}. PINNs incorporate the physical laws governing scientific data in the deep learning framework, resulting in improved accuracy and credibility.

As for the volumetric super-resolution of scientific data based on PINNs, MeshfreeFlowNet \cite{Jiang2020} builds a deep continuous spatio-temporal framework and takes physical constraints into account to achieve grid-free fluid super-resolution. TransFlowNet \cite{Wang2022} uses Transformer to remodel the encoder for more effective deep features and gives better super-resolution results. These approaches show that it is feasible and effective to consider physical constraints in a deep learning-based volumetric super-resolution framework. However, they all require high computational power, take a long training time, and prove difficult to converge. Furthermore, the massive sample points required in the training could generate memory-related problems. Other measures must be taken, such as reducing the batch size or the number of sample points, which may cause suboptimal training outcomes.

\subsection{Importance Sampling for Deep Learning}
Deep learning necessitates substantial data for training, leading to a significant increase in computing expenses. Nevertheless, it turns out that not all data or samples are equally important, and a considerable portion can be disregarded during training without affecting the final model. In recent years, researchers have begun to explore using importance sampling to accelerate the training of deep neural networks.

Alain et al. \cite{alain2015variance} presented a distributing neural network training method that uses a cluster of GPU workers to search for the most informative samples to train on. Chen et al. \cite{chen2018fastgcn} used importance sampling to reformulate the loss and the gradient for the fast learning of graph convolutional networks. Katharopoulos and Fleuret \cite{katharopoulos2018not} provided a practical upper bound to the gradient norm of any neural network and showed that not all samples are equal in the duration of training. By approximating the ideal sampling distribution using robust optimization, Johnson and Guestrin \cite{Johnson2018} proposed a practical importance sampling scheme for speeding up the training of deep learning models. Banerjee et al. \cite{banerjee2021deterministic} presented a selection algorithm to generate a deterministic sequence of mini-batches instead of a random one to train the deep neural networks.

These methods mainly focus on the training acceleration of deep neural networks via selecting training samples. In contrast, PINNs pay more attention to the distribution of sample points in the spatio-temporal dimension. There has also been some impressive work on importance sampling during PINNs training. Nabian et al. \cite{nabian2021efficient} sampled the collocation points based on a distribution proportional to the loss function, leading to an improvement in convergence performance when training PINNs. Lu et al. \cite{lu2021deepxde} proposed a residual-based adaptive refinement approach to improve the training efficiency of PINNs. This method repeatedly adds sample points in locations where the PDEs residual is large until the mean residual is smaller than a threshold. Yang et al. \cite{Yang2022} proposed dynamic mesh-based importance sampling to achieve stable convergence by estimating sample weights efficiently.

PINNs are designed for obtaining an approximate solution of the differential equation by training a neural network so that the fitting error is as small as possible. These importance sampling methods for PINNs are committed to evaluating the weight of each sample point, ignoring the dynamic overall importance distribution. In the physics-constrained volumetric super-resolution, the training data are cubes randomly cropped from the original data. The calculation of PDEs loss relies on the sampling within each cube. The distribution of sample points guides the gradient descent. Therefore, it is necessary to dynamically estimate the importance distribution of the original data during training.

For the deep learning-based volumetric super-resolution of scientific data, we propose an octree-based hierarchical sampling optimization to achieve importance sampling. It significantly improves the training performance by reducing the number of sample points.

\section{Method}\label{sec:method}
In this section, we propose a new hierarchical sampling method based on octree to perform multi-stage training. It focuses on training optimization and is mainly oriented to the volumetric super-resolution task based on physics-informed deep learning. Specifically, we use an octree-based data structure to achieve importance sampling on the latent context grid, which reduces the computation spent on the physics-constrained network. As a result, our method can greatly reduce the number of sample points while maintaining good convergence performance. Compared with other sampling methods applied to PINNs, our approach is elaborately designed for volumetric data and shows better stability.

\subsection{Overview}\label{sec:overview}
\begin{figure*}[htb]
\centering
\includegraphics[width=1.0\linewidth]{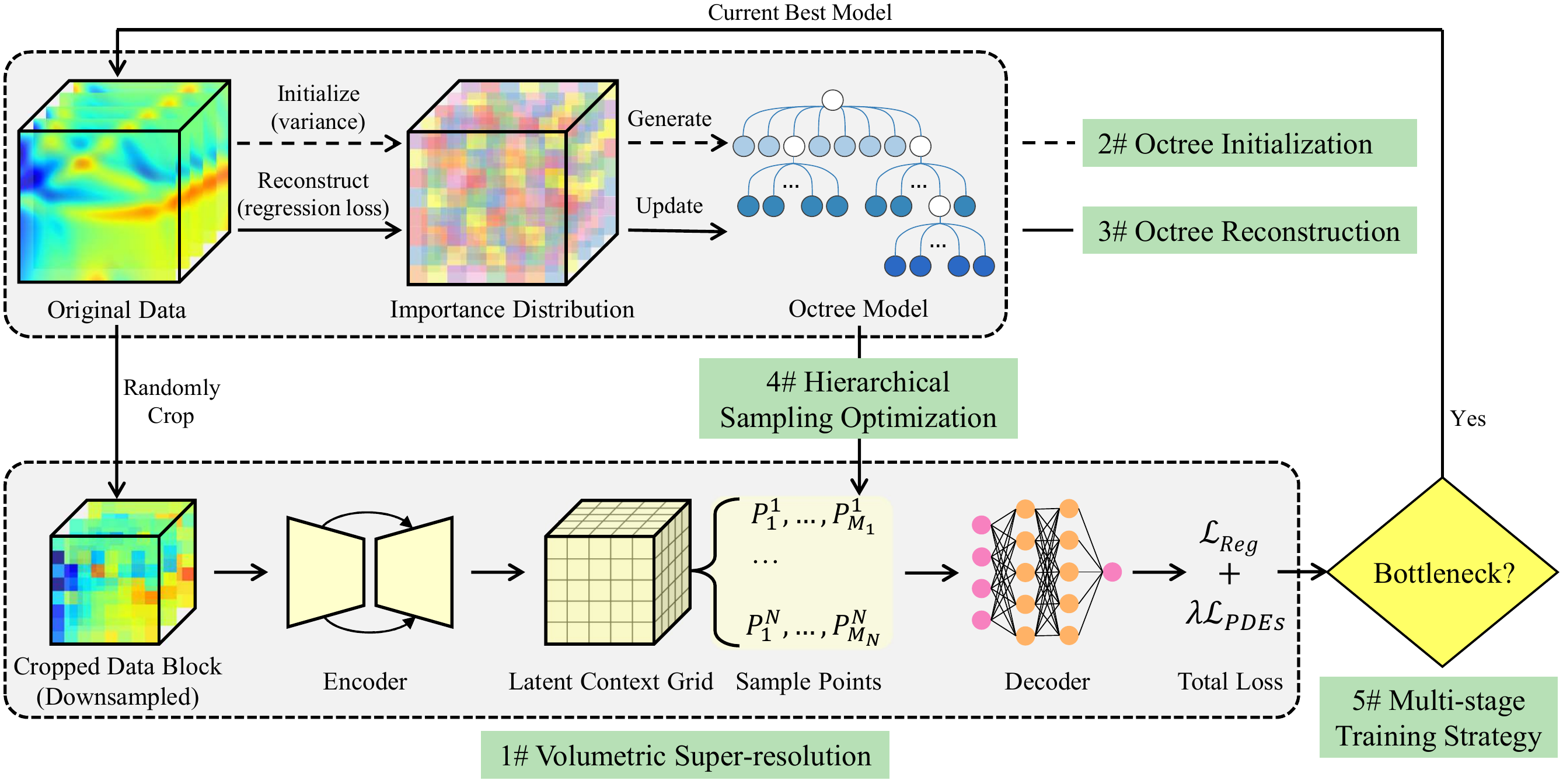}
\caption{The proposed octree-based hierarchical sampling optimization in the training pipeline. The octree model is initialized or reconstructed before each stage of training to achieve importance sampling by hierarchical sampling optimization. We number and highlight the five technical details corresponding to the subsections.}\label{fig:overview} 
\end{figure*}

The overview of our octree-based hierarchical sampling optimization in the training pipeline is shown in \autoref{fig:overview}. Before training, we take advantage of the variance information in the original data to generate an initial octree model. The octree subdivides the volumetric domain of data into several cuboid subblocks with different sizes. Each leaf node corresponds to an indivisible subblock. Then, the first stage of training begins. The volumetric super-resolution framework randomly crops cubic data blocks from the original data as ground truth and feeds them one by one into the encoder after downsampling. Based on the octree model, we perform hierarchical importance sampling on the latent context grid produced by the encoder. The number of sample points is adaptive to the training data block position in the original data. We compute the corresponding prediction at each sampling point through the decoder and estimate the average regression loss and PDEs loss for this training data block. After updating weights based on the total loss, we repeat the previous training process several times. When the first stage of training reaches a bottleneck, i.e., the loss is almost no longer decreasing, we reconstruct the octree based on the current best model. The updated octree ensures that more sample points are collected in areas with larger regression loss. We update the octree multiple times and after each update, a stage of training follows. Finally, the training ends when one of the termination conditions is reached.

We explain five technical details of the proposed method in the following subsections, including Volumetric Super-resolution, Octree Initialization, Octree Reconstruction, Hierarchical Sampling Optimization, and Multi-stage Training Strategy.

\subsection{Volumetric Super-resolution}\label{sec:volumetric}
Volumetric super-resolution refers to the generation of high-resolution spatio-temporal scientific data from low-resolution ones. Different from the interpolation methods, deep learning solutions use an encoder to extract implicit features of low-resolution data and a physics-constrained decoder to generate high-resolution data. The deep learning models work well but are difficult to train. Since the original data has a large dimensional size, several data blocks are randomly cropped to construct the training dataset.

We define the original data as $\mathbf{D}_{gt} \in \mathbb{R}^{h\times w\times t\times c}$, where $c$ is the number of physical variables. The cropped training data block is denoted as $\mathbf{D}_{gt}^b$. During training, $\mathbf{D}_{gt}^b$ is first downsampled to low-resolution $\mathbf{D}_{lr}^b$ and fed into the encoder to produce the latent context grid. Then, a large number of coordinates are randomly sampled in the latent context grid. Along with their corresponding features, these sample points are decoded to the estimated values of the original physical variables, which are used to compute the total loss. In one forward propagation, the encoder performs only once, while the decoder performs many times. Specifically, the decoder performs as many times as the number of sample points. Therefore, the number and distribution of sample points have a decisive impact on training efficiency. To this end, we improve the training performance of volumetric super-resolution solutions by octree-based hierarchical sampling optimization.

\subsection{Octree Initialization}\label{sec:initialization}
We use the original data to initialize the octree as preprocessing before training. The initialized octree will guide the importance sampling process on the latent context grid during the first stage of training. Specifically, we generate the first octree based on the variance of the original data to make the first stage of training pay more attention to areas with more significant numerical fluctuations. 

First, we compute the mean and variance of each physical variable by
\begin{gather}
    \label{equ:Ei-Vi}
        E_i = \frac{1}{n} \sum_{j=1}^n x_{ij} \\
        V_i = \frac{1}{n} \sum_{j=1}^n \left( x_{ij} - E_i \right)^2,
\end{gather}  
where $E_i$ and $V_i$ denote the mean and variance of the $i$-th physical variable, respectively. $n$ is the number of grid points in the volumetric data and equals $h\times w\times t$. $x_{ij}$ is the normalized value of the $i$-th physical variable at the $j$-th grid point. The volumetric super-resolution task for scientific data may require different weights for different physical variables. We define the total variance $V_{tot}$ of the volumetric data as
\begin{equation}
    \label{equ:Vtot}
        V_{tot} = \frac{1}{c} \sum_{i=1}^c \alpha_i V_i,
\end{equation}
where $c$ is the number of physical variables, and $\alpha_i$ denotes the custom weight of the $i$-th physical variable. The sum of $\alpha_i$ equals $c$. Obviously, $V_{tot}$ is a weighted sum of all $V_i$.

Then, we start to generate the first octree. Obviously, The first node of the octree is the root node that denotes the entire volumetric domain of data. As each dimension size of the volumetric data may be different, the volumetric domain is a cuboid. Next, we recursively produce leaf nodes from the root node. We first divide the cuboid into 8 cuboid subblocks of the same size, corresponding to producing 8 leaf nodes from the root node. After the first subdivision, we compute the variance $V_{tot}^{sub}$ for each subblock, in turn, using the same way as \autoref{equ:Vtot}. If a subblock meets $V_{tot}^{sub} \geq V_{tot}$ and one of its dimensions is greater than the custom size threshold $s$, we continue to subdivide this subblock. Specifically, we subdivide the subblock in three directions if all three dimensions are greater than $s$, in two directions if only two dimensions are greater, and so on. Correspondingly, each leaf node may produce 8, 4, or 2 leaf nodes. We perform the subdivision recursively until all the cuboid subblocks, i.e., the leaf nodes, do not meet the subdivision conditions. 

\begin{figure}[htb]
    \centering
        \subfloat[Volumetric domain]
        {\includegraphics[width=0.48\linewidth]{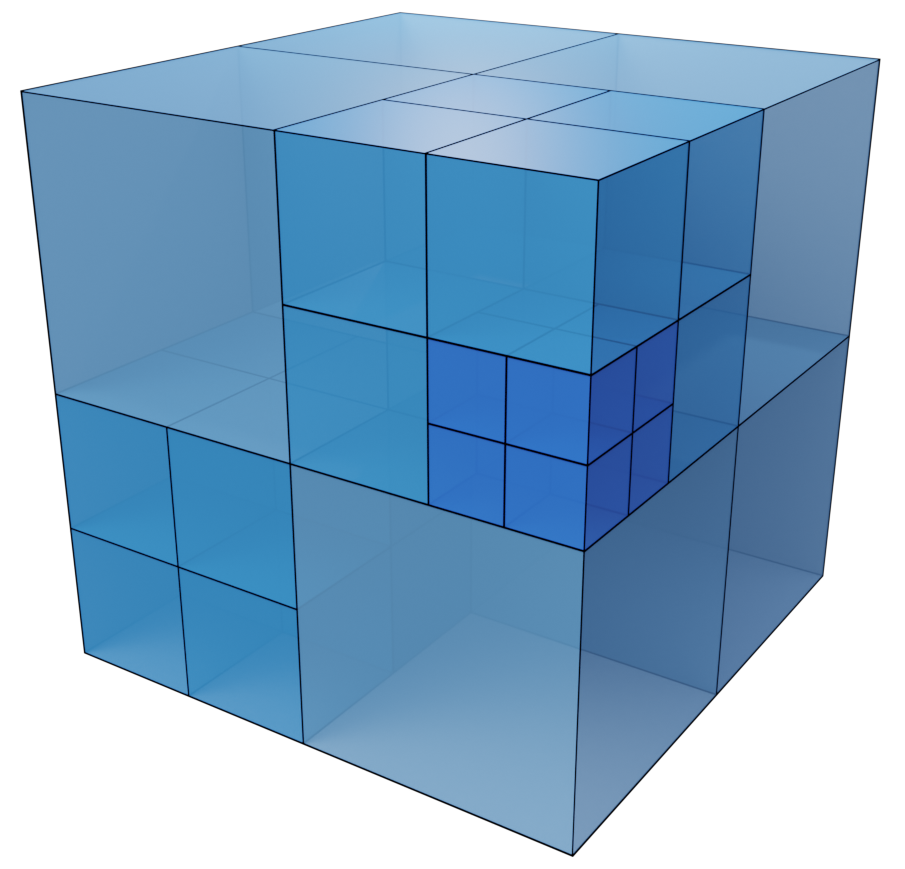}}\quad
        \subfloat[Octree stucture]
        {\includegraphics[width=0.48\linewidth]{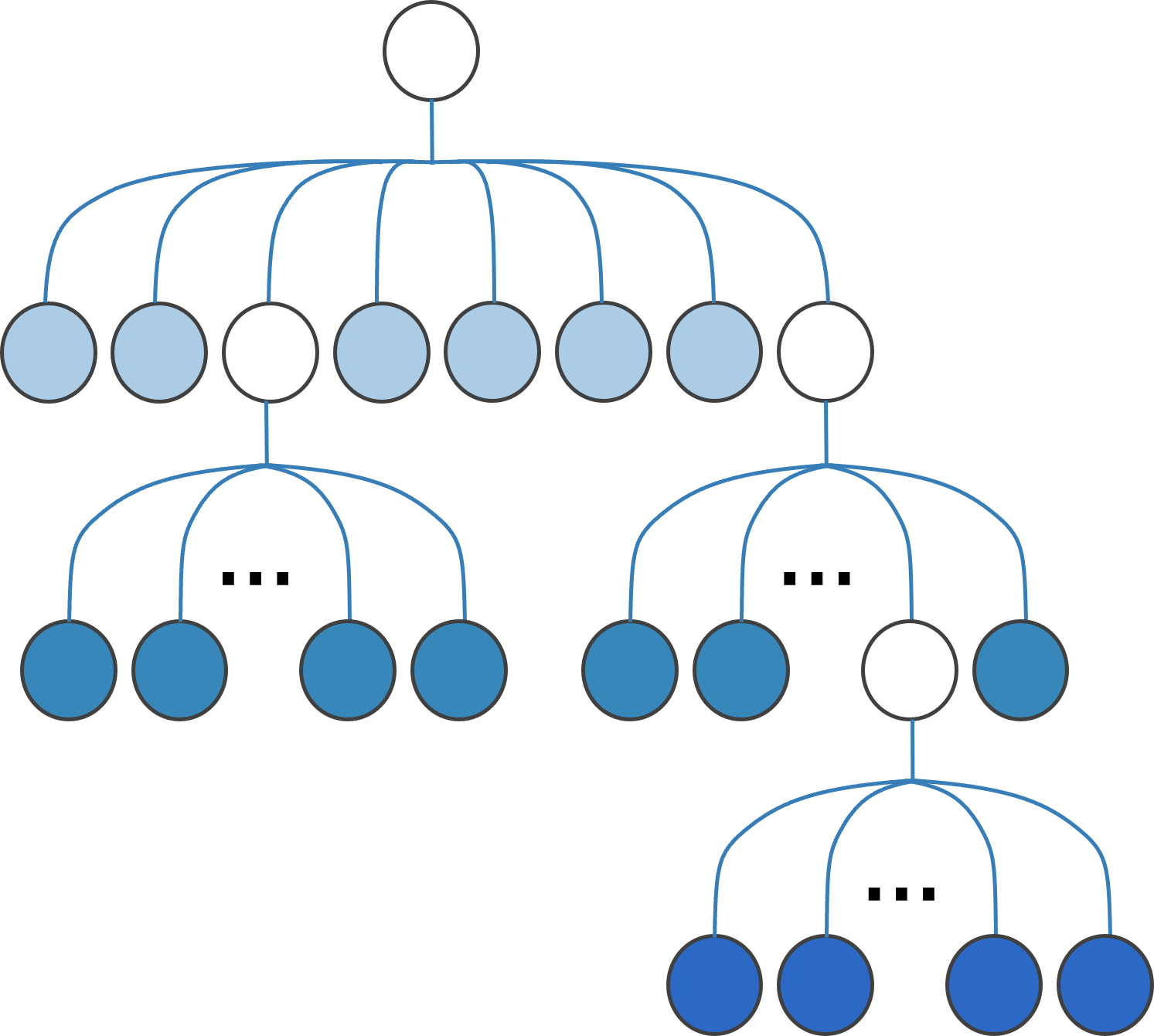}}
    \caption{\label{fig:octree}An octree model example. The root node of octree denotes the entire volumetric data, and each leaf node corresponds to an indivisible subblock. We mark leaf nodes in the same depth and their corresponding subblocks with identical colors.}	
\end{figure}

At last, we complete the octree initialization, and an octree model example is shown in \autoref{fig:octree}. The root node of octree denotes the entire volumetric data. Each leaf node corresponds to an indivisible subblock. The detailed octree subdivision algorithm is illustrated in \autoref{alg:subdivision}.

\begin{algorithm}[htb]
\KwIn{$\mathbf{D}_{gt}$, root, $V_{tot}$}
\SetKwProg{Fn}{Function}{:}{}
\SetKwFunction{Func}{Count}
\SetKwFunction{FMain}{Subdivide}

\Fn{\FMain{$\mathbf{D}_{gt}$, root, $V_{tot}$}}{

    \Fn{\Func{root}}{
      $n \longleftarrow 0$

        \lIf{root's height $\geq s_h$}{$n \longleftarrow n+1$}    
        \lIf{root's width $\geq s_w$}{$n \longleftarrow n+1$}
        \lIf{root's length $\geq s_l$}{$n \longleftarrow n+1$}
        
      \Return $2^n$
    }
    \textbf{End Function}
    
    Create \Func{root} nodes and add them to the root's leaves
    
    \ForEach{leaf $\in$ root's leaves}{
        Calculate $V_{tot}^{sub}$ for each leaf
        
        \uIf{$V_{tot}^{sub} \geq V_{tot}$}{\FMain{$\mathbf{D}_{gt}$, leaf, $V_{tot}$}}
    }
}
\textbf{End Function}
\caption{Octree Subdivision Algorithm}
\label{alg:subdivision}
\end{algorithm}

\subsection{Octree Reconstruction}\label{sec:reconstruction}
In the first stage of training, as the weights in the deep learning network are updated iteratively, the variance-based initial octree gradually fails to represent the distribution characteristics of training data under the current model. This situation leads to importance sampling gradually playing an irrelevant role in training, and the descending gradient of the total loss will converge to 0. Therefore, we propose a loss-based octree reconstruction scheme. The reconstructed octree represents the regression loss distribution on training data, and further plays the significant role of importance sampling to ensure the convergence and effectiveness of subsequent stages of training.

Since the inference is time-consuming using the original data $\mathbf{D}_{gt}$, we downsample it using a scale factor $a$ to obtain the Ground Truth data $\mathbf{D}_{gt}^a$ for octree reconstruction. This pre-processing will reduce the inference time to $\frac{1}{a^3}$. It is a compromise between the accuracy of the importance distribution and the speed of octree reconstruction.

First, we perform inference using $\mathbf{D}_{gt}^a$ and the current best model. The high-resolution result is
\begin{equation}
    \label{equ:D-hr}
    \mathbf{D}_{hr}^k = \mathcal{N}_{best}^k \left( \mathbf{D}_{lr}^a \right),
\end{equation}
where $\mathbf{D}_{lr}^a$ is the low-resolution input after downsampling $\mathbf{D}_{gt}^a$. $\mathcal{N}_{best}^k(\cdot)$ is the best model obtained after the $k$-th stage of training and $\mathbf{D}_{hr}^k$ is the high-resolution output after inference.

Then, we use $\mathbf{D}_{hr}^k$ and $\mathbf{D}_{gt}^a$ to compute the regression loss distribution. The regression loss at the $j$-th grid point is defined as
\begin{equation}
    \label{equ:L-j-k}
        L_j^k = \frac{1}{c} \sum_{i=1}^c \alpha_i \Vert x_{ij} - \hat{x}_{ij}^k \Vert_1 
        ,
\end{equation}
where $\hat{x}_{ij}^k$ is the estimated value of the $i$-th physical variable at the $j$-th grid point, which can be obtained by indexing in $\mathbf{D}_{hr}^k$. The total regression loss is
\begin{equation}
    \label{equ:L-tot-k}
        L_{tot}^k = \frac{1}{n} \sum_{j=1}^n L_j^k.
\end{equation}

Finally, similar to the initial octree, we recursively reconstruct the octree from the root node based on the regression loss. We denote the regression loss of a subblock by $L_{tot}^{sub,k}$, which can be obtained in the same way as \autoref{equ:L-tot-k}. The subdivision conditions are $L_{tot}^{sub,k} \geq L_{tot}^k$ and there is a dimension greater than $s$ in the subblock.

\subsection{Hierarchical Sampling Optimization}\label{sec:sampling}
As described in \autoref{sec:volumetric}, after randomly cropping a cubic data block from the original data as the ground truth for a single training cycle, the encoder of the volumetric super-resolution framework will extract features from the downsampled low-resolution data block and build a latent feature grid. Then, most methods randomly or uniformly take sample points from the latent context grid and feed them to the decoder to estimate the values of physical variables at the corresponding positions. Since the training performance is significantly affected by the sample points, we present an octree-based hierarchical sampling method to optimize the distribution of sample points in each training data block.

As shown in \autoref{fig:octree}, each leaf node in the octree corresponds to an indivisible cuboid subblock in the original data. We define the number of sample points in each subblock with the same value $m$. Each leaf node may have a different depth, so each subblock may have a different size. This setup leads to different sample point densities in different subblocks, which is the key to achieving importance sampling.

\begin{figure}[htb]
    \centering
    \includegraphics[width=1.0\linewidth]{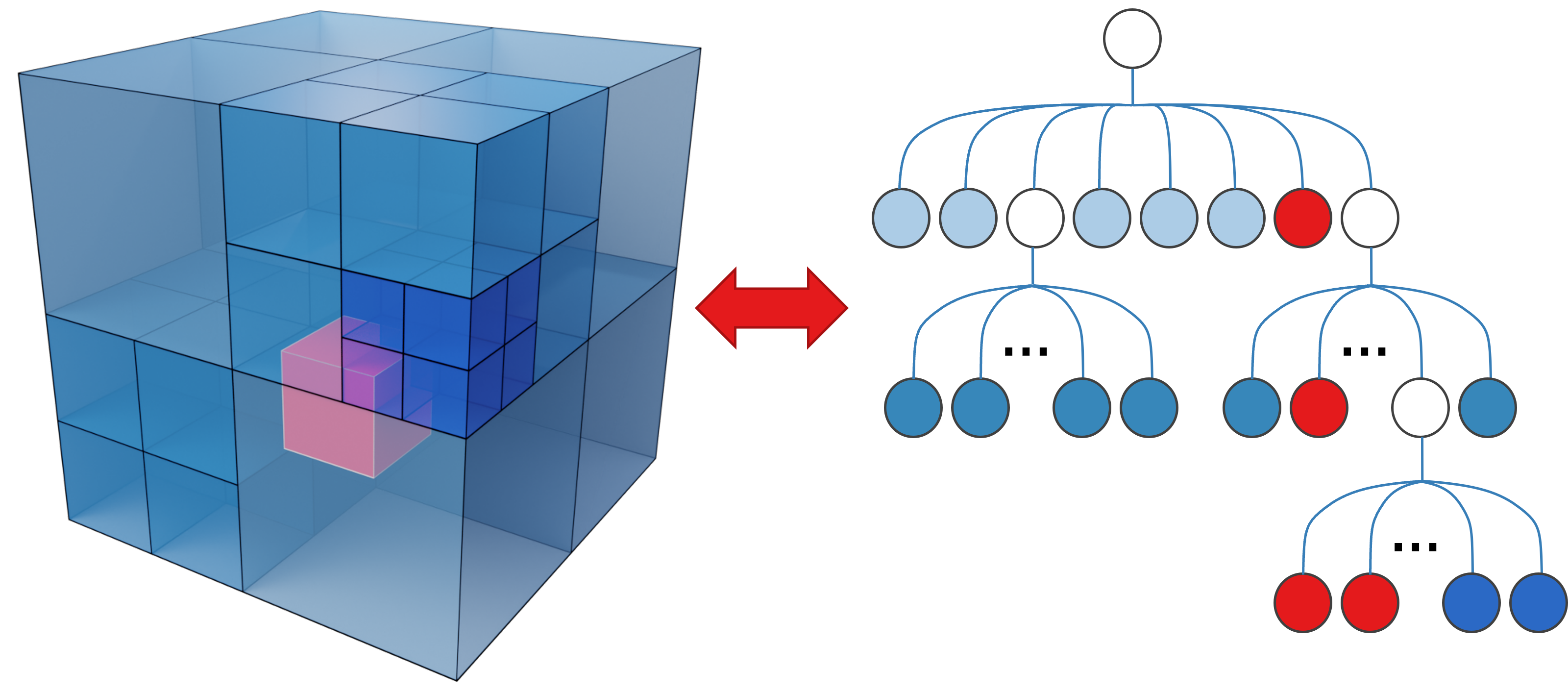}
    \caption{\label{fig:sampling}The schematic diagram showcases the octree-based hierarchical sampling optimization. The red cube indicates the position of the training data block in the volumetric domain (left), and the red leaf nodes indicate the subblocks intersecting with the training data block (right).}	
\end{figure}

In a randomly cropped training data block, the implicit feature of a sample point can be indexed in the latent context grid by its position, so the sampling process is to determine the position distribution of sample points. The schematic diagram of octree-based hierarchical sampling optimization is shown in \autoref{fig:sampling}. 

First, based on the size and position of the training data block, we retrieve the leaf nodes that intersect with it and get several intersection spaces. These intersection spaces correspond to the red leaf nodes in the right part of \autoref{fig:sampling}.

Then, we determine the number of sample points in each intersection space by
\begin{equation}
    \label{equ:M}
        M = \lceil m\frac{S_{int}}{S_{leaf}} \rceil,
\end{equation}
where $S_{int}$ and $S_{leaf}$ are the volumes of the intersection space and its corresponding leaf node, respectively. When generating the octree, we can easily obtain $S_{leaf}$ based on the size of each indivisible cuboid subblock. Together with its position and the size and position of the training data block, we can also compute $S_{int}$.

Finally, we perform random sampling in each intersection space in turn, and the number of sample points is computed by \autoref{equ:M}. Since the leaf nodes corresponding to the intersection spaces may have different depths, the sampling process on the training data block is hierarchical. As a result, we achieve the octree-based importance sampling because of the different sampling densities within each intersection space.

\subsection{Multi-stage Training Strategy}\label{sec:training}
As described in \autoref{sec:reconstruction}, the octree-based importance sampling will gradually fail as the training proceeds and the weights update. Therefore, we perform the training process with several stages, called the multi-stage training strategy. Whenever reaching a bottleneck, we terminate the current stage of training and reconstruct the octree to start the next stage of training. At each stage of training, we perform hierarchical sampling optimization using the updated octree.

The key to multi-stage training is when to terminate the current stage and start the next one. In order to avoid frequently starting a new stage and performing the time-consuming octree reconstruction, we define $q$ as the minimum number of epochs for each stage of training. As we all know, if the loss is no longer decreasing and has not converged to the expected value, we call the training reaching a bottleneck. Therefore, starting from the $q$-th epoch ($q \geq 10$), we use loss values of the last 10 epochs to do linear regression. If the mean value is less than or equal to threshold $Q$, we terminate the training immediately and save the current best model as the training result. If the mean value is higher than threshold $Q$ and the slope of the linear regression model is greater than or equal to 0, we start the next stage of training.

\section{Experiments}\label{sec:experiments}
There are two recently published state-of-the-art solutions of spatio-temporal super-resolution for scientific data, called MeshfreeFlowNet \cite{Jiang2020} and TransflowNet \cite{Wang2022}, respectively. We apply the proposed octree-based hierarchical sampling optimization to these two deep-learning methods and verify its effectiveness with extensive experiments on the Rayleigh-B{\'e}nard convection problem. We first evaluate the performance of our sampling optimization method in terms of both convergence speed and model accuracy. Then we perform an ablation study to elucidate the role of the loss-based octree reconstruction scheme and discuss the optimal configuration of custom parameters. At last, we experiment with our method in the application of solving PDEs using PINNs and compare it with other importance sampling methods like DeepXDE \cite{lu2021deepxde}. The results show that the octree-based hierarchical sampling optimization is an effective technology for importance sampling of volumetric data, and it can be generalized and applied in PINNs-based solutions. In addition, the experiments are implemented on a high-performance computer with one NVIDIA GeForce RTX 3090.

\subsection{Baseline}\label{sec:baseline}
MeshfreeFlowNet is a deep continuous spatio-temporal super-resolution framework with PDEs loss, and TransFlowNet is a physics-constrained volumetric super-resolution solution with Transformer blocks. They focus on solving the high-resolution reconstruction problem of 3D flow simulations using deep learning technology. In the training phase of both frameworks, multiple dense sampling of the latent context grid is required. The distribution of sample points significantly impacts the training performance and results. We use them as baselines and apply our octree-based hierarchical sampling optimization in their training phase. 

The specific training settings of baselines are detailed as follows. We train the original MeshfreeFlowNet and TransFlowNet using an Adam optimizer with a learning rate 0.001 for 100 epochs. In each epoch, we take 3000 randomly cropped data blocks as training samples. Limited by the memory size, the batch size is set to 10. The weighting coefficient of PDEs loss is 0.02. Besides, we set the number of sample points in the latent context grid the same as the original papers to 512.

\subsection{Rayleigh-B{\'e}nard Convection}\label{sec:RBProblem}
\textbf{Problem description.} Rayleigh-B{\'e}nard convection is a model for turbulent convection of a fluid confined between two thermally conducting plates. The upper and lower plates are cold and hot respectively. The fluid particles tend to produce vortices and turbulent regiments due to a temperature gradient. %Its governing PDEs are given by
We are concerned with four physical variables, which are pressure $p$, temperature $T$, velocity component $u$ in the $x$ direction, and velocity component $w$ in the $z$ direction. In this section, we use the same setup with MeshfreeFlowNet to run the flow simulation.

\textbf{Dataset generation.} We use Dedalus framework \cite{Burns2020Dedalus} to numerically solve the governing PDEs of Rayleigh-B{\'e}nard convection problem. The obtained Ground Truth data has spatio-temporal dimensions of $128\times 512\times 200$ $(h\times w\times t)$. During training, we continuously crop some data blocks of $128\times 128\times 16$ to constitute the training samples for each epoch. The cropped data blocks are downsampled to $16\times 16\times 4$ as the low-resolution input of deep learning models. We also generate another Ground Truth data for the evaluation by using different initial conditions and boundaries.

\textbf{Training details.} The training settings of baselines are explained in \autoref{sec:baseline}. We use the same setup for our octree-based sampling optimization method. In addition, we list the values of all hyper-parameters in \autoref{tab:hyper-parameters}.

\begin{table}[htb]
    \centering
    \caption{\label{tab:hyper-parameters}Hyper-parameters used for Rayleigh-B{\'e}nard convection problem. $\alpha_i$ is the weight of the $i$-th physical variable, $s$ is the dimension size threshold, $a$ is the scale factor for  octree reconstruction, $m$ is the number of sample points in each indivisible subblock, $q$ is the number of epochs to evaluate the training stage, and $Q$ is the loss threshold.}
    \resizebox{\linewidth}{!}{%
        \centering
        \begin{tabular}{ccccccccc}
        \toprule
        Parameter & $\alpha_i$ & $s_h$ & $s_w$ & $s_l$ & $a$ & $m$ & $q$ & $Q$ \\
        \midrule
        Value & 1 & 128 & 128 & 16 & 2 & 64 & 10 & 0.02 \\  
        \bottomrule
        \end{tabular}
    }
\end{table}

\textbf{Evaluation metrics.} To compare training performance and convergence behavior, we take snapshots of the deep learning models as they reach a specific level of convergence. We report the number of iteration epochs and the training time, denoted as $S$ and $T$, respectively. The number of sample points is defined as $N$. We also provide convergence curves of 100 epochs to show the detailed convergence process intuitively. Besides, we use peak signal to noise ratio (PSNR) and structural similarity index (SSIM) to quantitatively evaluate the prediction accuracy. We compute the average of the weighted physical variables as the final results. We also report the total training time and the total number of sample points of 100 epochs, denoted as $T_{100}$ and $N_{100}$, respectively. The total number of octree reconstructions during training is defined as $N_{re}$. In each table and figure, w/o is the original version of the baseline, and w is the optimized version using our octree-based sampling method.

\begin{figure}[!h]
\centering
    \subfloat[MeshfreeFlowNet\cite{Jiang2020}]{\includegraphics[width=0.8\linewidth]{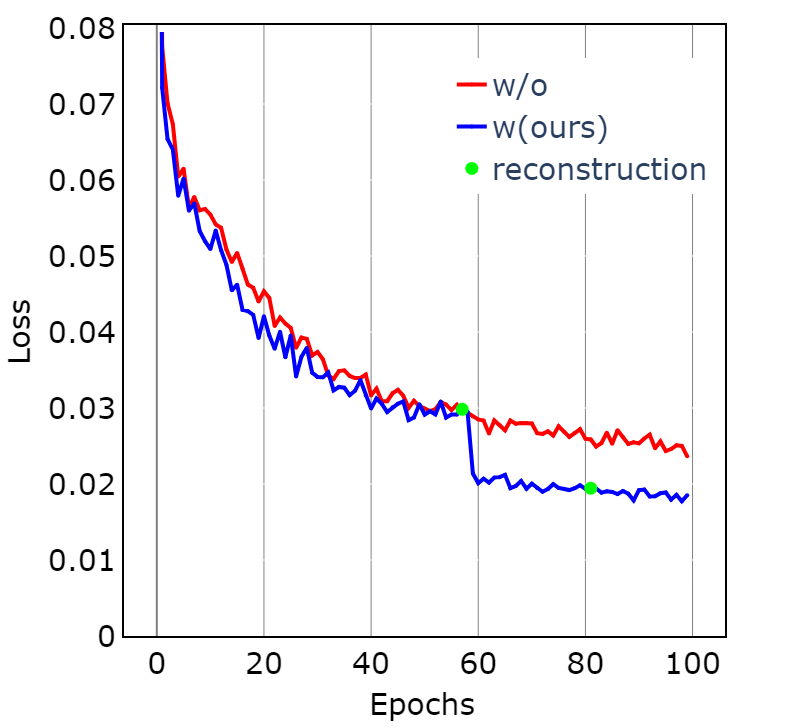}} \quad
    %\vspace{0.02\linewidth}
    \subfloat[TransFlowNet\cite{Wang2022}]{\includegraphics[width=0.8\linewidth]{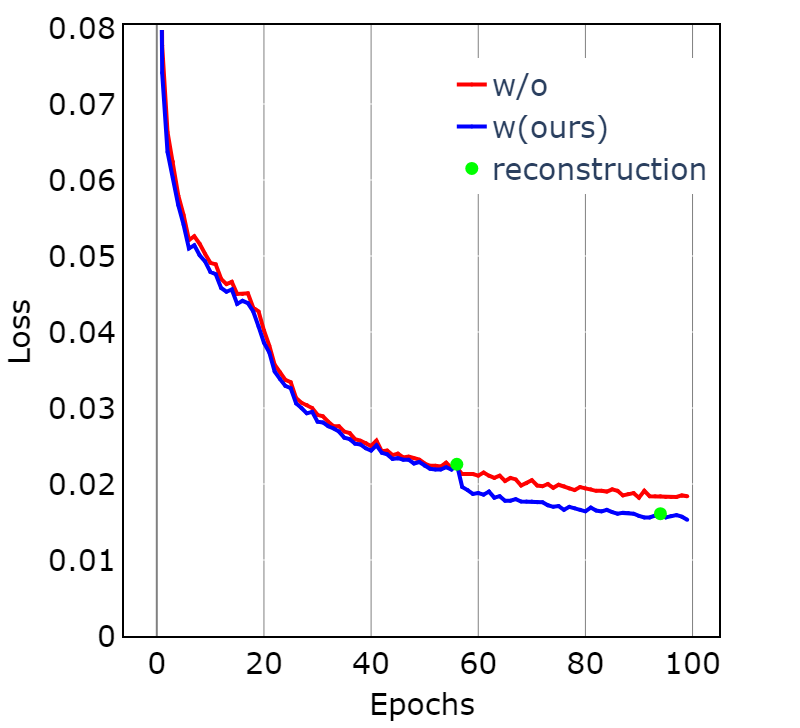}}
    \caption{\label{fig:RB_curves}The convergence curves of MeshfreeFlowNet and TransFlowNet, respectively. The solid green circles mark the places where the octree reconstruction is performed.}	
\end{figure}

\textbf{Results comparison.} \autoref{tab:RB_train} shows the quantitative evaluation of training performance on Rayleigh-B{\'e}nard convection problem. It summarizes the number of iteration epochs $S$, the training time $T$, and the number of sample points $N$, at the convergence level of 0.04. Benefiting from importance sampling, our method achieves a significant reduction of sampling points with the same convergence level ($57.14\%$ reduction in MeshfreeFlowNet and $42.50\%$ reduction in TransFlowNet). As a result, the training time is $1.67 \times$ faster in MeshfreeFlowNet and $1.32\times$ faster in TransFlowNet. Our method also requires fewer epochs to achieve a convergence level consistent with the baselines. Besides, as shown in \autoref{fig:RB_curves}, the convergence curves indicate that our octree-based sampling optimization method can help the training loss descend more efficiently. Although the convergence acceleration of our method is not noticeable at first, the reconstructed octree improves the convergence performance by leaps and bounds, especially in MeshfreeFlowNet. Our method eventually provides a lower training loss in the same number of iterations. 

\begin{table}[htb]
    \centering
    \caption{\label{tab:RB_train}Quantitative evaluation of training performance on Rayleigh-B{\'e}nard convection problem. The convergence level is 0.04.}
    \resizebox{\linewidth}{!}{%
        \centering
        \begin{tabular}{ccccc}
        \toprule
        \multirow{2}*{Method} & \multicolumn{2}{c}{MeshfreeFlowNet\cite{Jiang2020}} & \multicolumn{2}{c}{TransFlowNet\cite{Wang2022}} \\
        \cmidrule(lr){2-3}\cmidrule(lr){4-5}
        & w/o & \textbf{w(ours)} & w/o & \textbf{w(ours)} \\
        \midrule
        $S$        & 27   & \textbf{20}   & 22   & \textbf{21}   \\
        $T$/min    & 45   & \textbf{27}   & 84   & \textbf{49}   \\
        $N$/$10^8$ & 0.28 & \textbf{0.12} & 0.34 & \textbf{0.19} \\
        \bottomrule
        \end{tabular}
    }
\end{table}

\begin{figure}[!h]
\centering
    \subfloat[MeshfreeFlowNet\cite{Jiang2020}]{\includegraphics[width=1.0\linewidth]{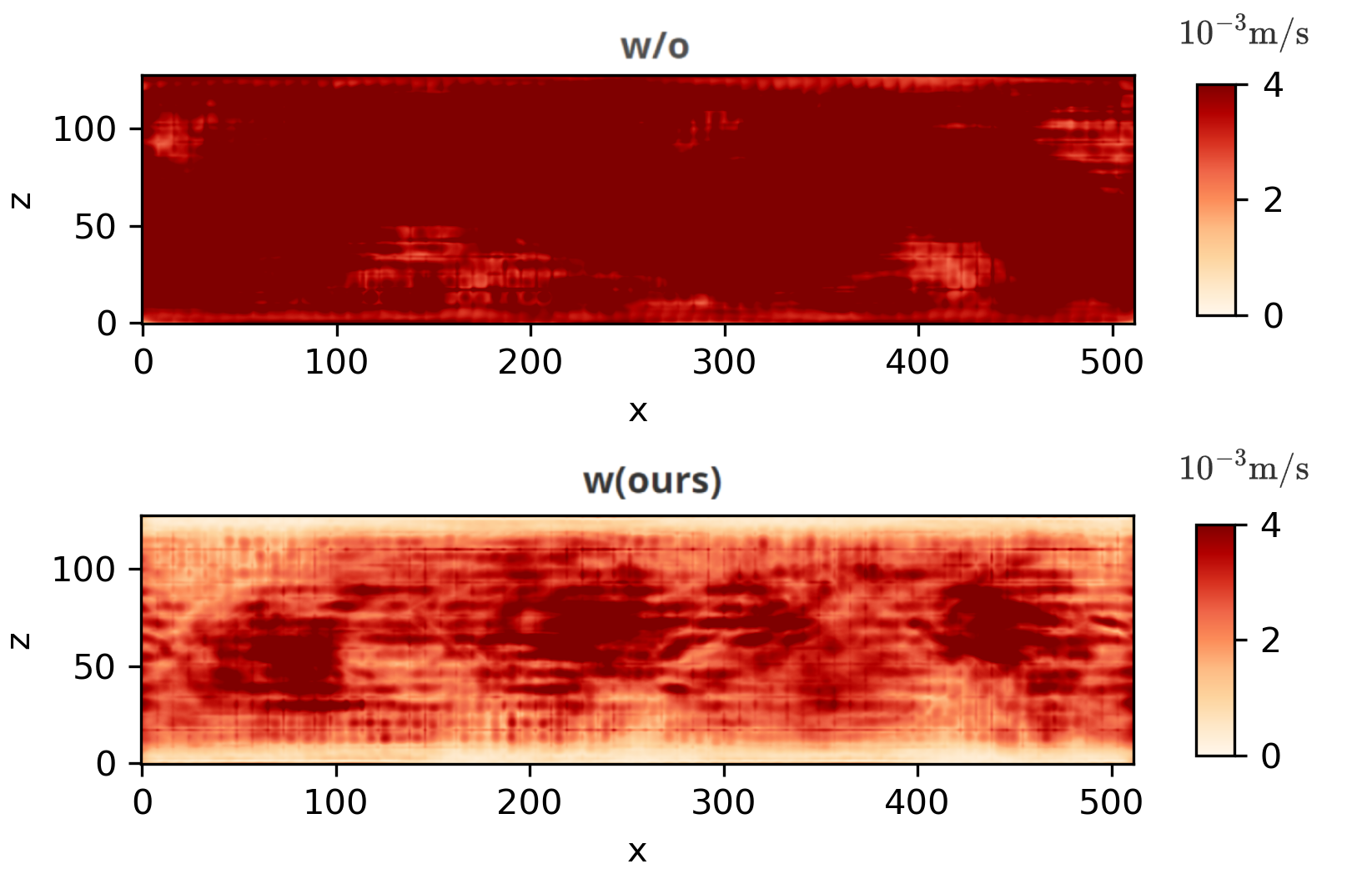}}\quad
    %\vspace{0.02\linewidth}
    \subfloat[TransFlowNet\cite{Wang2022}]{\includegraphics[width=1.0\linewidth]{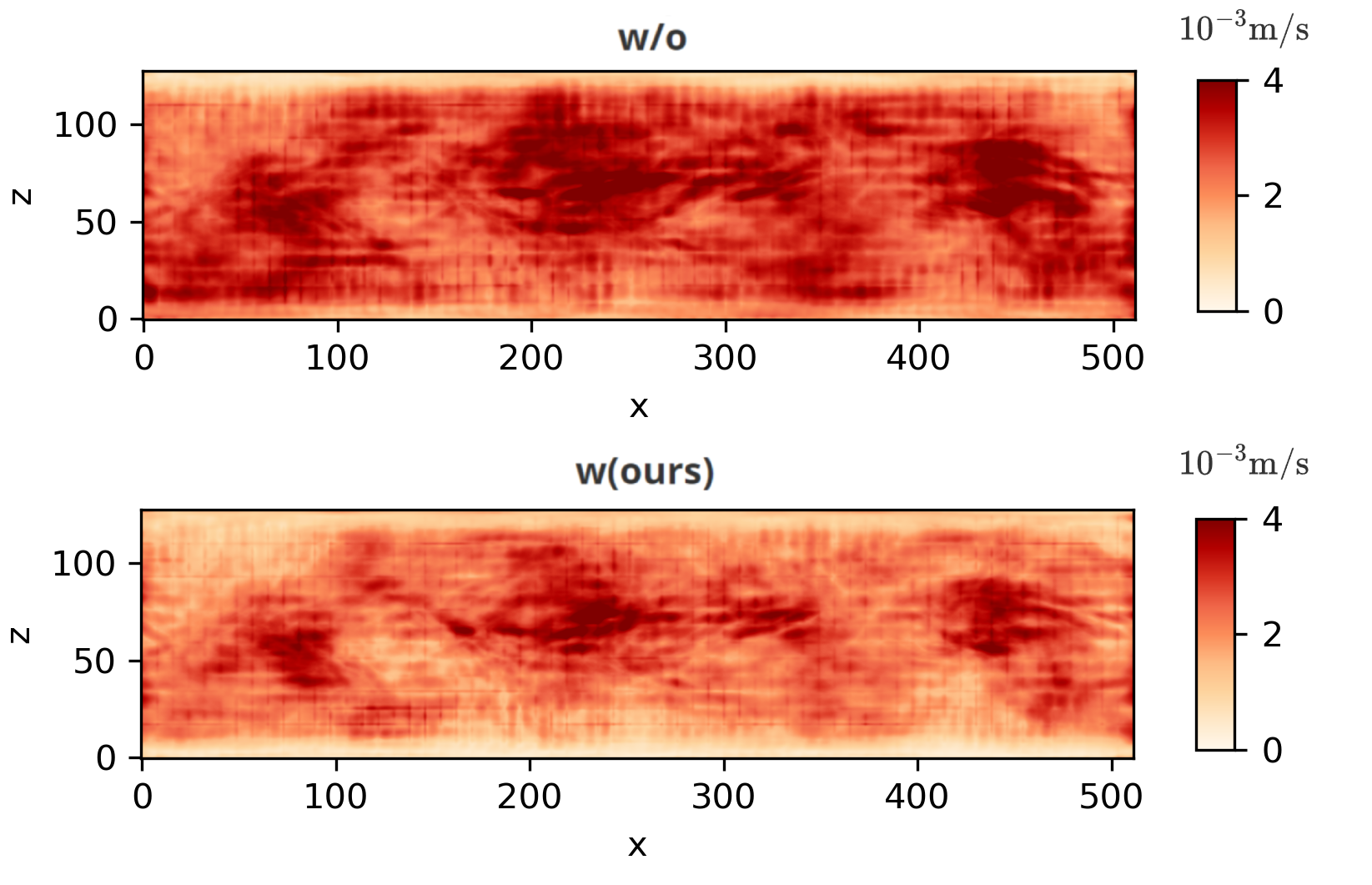}}
    \caption{\label{fig:RB_errors}The MAE of MeshfreeFlowNet and TransFlowNet at each position in the $x$-$z$ space, respectively. We use the velocity component $u$ in the $x$ direction as an example.}	
\end{figure}

\autoref{tab:RB_accuracy} presents the quantitative comparison of prediction accuracy. For each method, we use the best model after training 100 epochs to perform the evaluation. Compared to baselines, the results of our octree-based sampling method show improvements in both PSNR and SSIM. Not only that, our method uses fewer sample points and shorter training time to produce better models. In MeshfreeFlowNet, our sampling optimization makes the number of sample points reduce by $37.25\%$, and the training accelerated by $17.58\%$. In TransFlowNet, they are $36.36\%$ and $20.00\%$, respectively. As the time-consuming octree reconstruction performs twice, the improvement in training time is not as significant as the number of samples. We also provide intuitive comparisons by visualizing the mean absolute error (MAE) of velocity component $u$ at each position in the $x$-$z$ space. The MAE at each position is an average of the absolute error at all moments. As shown in \autoref{fig:RB_errors}, our method gives fewer wrong predictions and significantly improves the results, especially in MeshfreeFlowNet.

\begin{table}[htb]
    \centering
    \caption{\label{tab:RB_accuracy}Quantitative comparison of prediction accuracy on Rayleigh-B{\'e}nard convection problem. The number of training epochs is 100.}
    \resizebox{\linewidth}{!}{%
        \centering
        \begin{tabular}{ccccc}
        \toprule
        \multirow{2}*{Method} & \multicolumn{2}{c}{MeshfreeFlowNet\cite{Jiang2020}} & \multicolumn{2}{c}{TransFlowNet\cite{Wang2022}} \\
        \cmidrule(lr){2-3}\cmidrule(lr){4-5}
        & w/o & \textbf{w(ours)} & w/o & \textbf{w(ours)} \\
        \midrule
        PSNR    & 52.03 & \textbf{53.21} & 52.71 & \textbf{53.74} \\
        SSIM    & 0.9894 & \textbf{0.9954} & 0.9952 & \textbf{0.9953} \\
        $T_{100}$/min    & 165   & \textbf{136}   & 386   & \textbf{237}   \\   
        $N_{100}$/$10^8$ & 1.02 & \textbf{0.64} & 1.54 & \textbf{0.96} \\
        $N_{re}$   & - & 2 & - & 2 \\
        \bottomrule
        \end{tabular}
    }
\end{table}

\subsection{Ablation Study}\label{sec:ablationstudy}
To further demonstrate the effectiveness of our octree-based sampling optimization approach, we investigate a variety of octree reconstruction strategies. Specifically, We study the impact of hyper-parameter $q$ using the Rayleigh-B{\'e}nard convection dataset and MeshfreeFlowNet. 

\begin{table*}[htb]
    \centering
    \caption{\label{tab:ablation}Ablation studies of the octree reconstruction using the Rayleigh-B{\'e}nard convection dataset and MeshfreeFlowNet. We first compare the training performance at the convergence level of 0.04. Then, we use the best model after training 100 epochs to perform the prediction evaluation. w/o is the original version of MeshfreeFlowNet, and the others are different designs of the octree reconstruction strategy.}
    \resizebox{0.7\linewidth}{!}{%
        \centering
        \begin{tabular}{ccccccccccccc}
        \toprule
        Design & w/o\cite{Jiang2020} & $q = 0$ & \textbf{$q = 5$} & $q = 10$ & $q = 15$ & $q = 30$ & $q = +\infty$\\
        \midrule
        $S$        & 27  & 23   & 23  & 20  & 22  & 22  & 20 \\
        $T$/min    & 45   & 78    & 32  & 27  & 29   & 30   & 27\\
        $N$/$10^8$ & 0.28 & 0.16  & 0.14 & 0.12 & 0.13 & 0.13   & 0.12 \\
        \midrule
        PSNR     & 52.03  & 52.10  & 52.05 & \textbf{53.21}  & 52.91  & 52.71  & 51.72  \\
        SSIM     & 0.9894  & 0.9891  & 0.9920 & \textbf{0.9954}  & 0.9930  & 0.9952  & 0.9927  \\
        $T_{100}$/min    & 165   & 339   & 149  & 136  & 131 & 120 & 118  \\
        $N_{100}$/$10^8$ & 1.02 & 0.65 & 0.65  & 0.64  & 0.61 & 0.63 & 0.60 \\
        $N_{re}$ & -      & 99   & 9  & 2  & 2   & 1   & 0   \\
        \bottomrule
        \end{tabular}
    }
\end{table*}

In different designs of MeshfreeFlowNet, w/o is the original version without sampling optimization, and different values of $q$ indicate different octree reconstruction strategies. $q = 0$ means that once the training of an epoch is finished, we update the octree and start the next stage of training immediately without bottleneck evaluation. $q = 5$ means we start bottleneck evaluation from the 5th epoch and correspondingly use loss values of the last 5 epochs to do linear regression. We use the multi-stage training strategy described in \autoref{sec:training} for $q \geq 10$. Besides, $q = +\infty$ means that we never reconstruct the octree and use the initialized octree forever. 

We use the eight evaluation metrics in \autoref{sec:RBProblem} to compare different octree reconstruction strategies on training performance and results. We first collect training data at the convergence level of 0.04 to evaluate training performance. Then, we compare the predictive performance with the best model after training for 100 epochs. We also provide the total training time and the total number of sample points of 100 epochs. We summarize all the results in \autoref{tab:ablation}.

As shown in the training performance lines of \autoref{tab:ablation}, when the octree-based sampling optimization method is introduced, the number of sample points $N$ is significantly reduced. At the same time, the number of iteration epochs $S$ at the convergence level of 0.04 is slightly improved. The training time $T$ of $q = 0$ is longer than w/o because octree reconstruction is performed after each training epoch. It takes more time than sample point reduction saves. In addition, since only the initial octree is used before reaching the convergence of 0.04, the training times of $q \geq 5$ are very close. The slight difference in training time between different designs results from random sampling within each intersection space.

As shown in the prediction performance lines of \autoref{tab:ablation}, our octree-based sampling optimization method greatly reduces the total number of sampling points $N_{100}$ for training 100 epochs. The PSNR and SSIM of $q = 0$ are nearly the same with w/o because frequent octree reconstructions fail to achieve importance sampling. It also increases the training time by $105.45\%$ due to the time-consuming octree reconstructions, although the number of sample points is reduced by $36.27\%$. The prediction performance of $q = +\infty$ has no promotion either. As described in \autoref{sec:reconstruction}, the reason is the variance-based initial octree gradually fails to represent the importance distribution of training data. Since octree is never reconstructed during training, it accelerates the training by $28.48\%$ with a reduction of sample points by $41.18\%$. The number of octree reconstructions $N_{re}$ of $q = 5$ is 9, leading to a minor improvement of PSNR, SSIM, and the total training time $T_{100}$. It indicates the effectiveness of a suitable octree reconstruction strategy in rebuilding the importance distribution of training data. 

The number of reconstructions should be determined with caution. We find a balance between the training time and the prediction accuracy with $N_{re} = 2$ when training 100 epochs. The $q = 10$ design gives the best results in PSNR and SSIM.

\subsection{Octree-based Sampling Optimization for PINNs}\label{sec:PINNs}
Although our method is designed for the volumetric super-resolution of scientific data, it can also be used in training tasks of other deep neural networks. We further explore the feasibility of the octree-based sampling optimization approach for the training of PINNs.

We use PINNs for solving the inverse problem of PDEs as an example. The inverse problem of PDEs aims to estimate unknown parameters in equations based on limited training data. It combines regression loss and PDEs loss as the total loss for training. It is worth noting that we only have training data in a specific area and PDEs with unknown parameters before training PINNs. Unlike the volumetric super-resolution of scientific data, the Ground Truth data in the whole spatio-temporal grid is unknown. Therefore, the sampling process only takes place in the specific area, as does the octree initialization and octree reconstruction. It means there is no need to compute the intersection spaces described in the \autoref{sec:sampling}. Using the octree generated for that specific area, we can perform sampling sequentially in each indivisible subblock that corresponds to each leaf node.

\textbf{Baseline.} We take the original PINNs \cite{Raissi2019PINN} and DeepXDE \cite{lu2021deepxde} as baselines. The original PINNs perform random sampling throughout training. The DeepXDE uses a residual-based adaptive refinement approach to improve the training efficiency. It achieves importance sampling by sampling more points near the ones that produced a bigger loss in the last forward propagation.

\textbf{Navier-Stokes equations.} Navier-Stokes equations are the PDEs that describe the flow of incompressible fluids in fluid mechanics. The 2D Navier-Stokes equations with unknown parameters are given explicitly by

\begin{gather}
    \label{equ:ns}
        \frac{\partial u}{\partial t} + \lambda_1(u\frac{\partial u}{\partial x} + v\frac{\partial u}{\partial y}) = -\frac{\partial p}{\partial x} + \lambda_2(\frac{\partial^2 u}{\partial x^2} + \frac{\partial^2 u}{\partial y^2}) \\
        \frac{\partial v}{\partial t} + \lambda_1(u\frac{\partial v}{\partial x} + v\frac{\partial v}{\partial y}) = -\frac{\partial p}{\partial y} + \lambda_2(\frac{\partial^2 v}{\partial x^2} + \frac{\partial^2 v}{\partial y^2}) \\
        \frac{\partial u}{\partial x} + \frac{\partial v}{\partial y} = 0,
\end{gather} 
where $u$ denotes the velocity component in the $x$ direction, $v$ the velocity component in the $y$ direction, and $p$ the pressure. $\lambda_1$ and $\lambda_2$ are the unknown parameters. PINNs are used to learn the parameters $\lambda_1$ and $\lambda_2$ that best describe the observed data in a specific area, as well as the pressure $p$.

We consider the problem that incompressible flow passes a circular cylinder, which is detailed described in \cite{Raissi2019PINN}. $\lambda_1$ and $\lambda_2$ are set to 1 and 0.01, respectively. We use the same high-resolution data set solved by NekTar\cite{karniadakis2005spectral}. 

\textbf{Implementation details.} We initialize the first octree based on the limited training data in a specific area of the cylinder wake. With the help of the first octree, we train PINNs by importance sampling of input coordinates from the beginning. When total loss reaches the bottleneck, we compute the loss value at each coordinate in the spatio-temporal grid according to the current best model. Then, we rebuild the octree based on the loss distribution. Similar to \autoref{sec:reconstruction}, we reconstruct the octree whenever the total loss reaches a bottleneck, avoiding failure of the importance sampling. All sampling operations are always performed in the entire domain by sampling in each indivisible subblock sequentially.

\textbf{Results comparison.} We use the L2 relative error between the predicted values and the exact solutions to evaluate the accuracy of models. The results are summarized in \autoref{tab:PINNs}.

\begin{table}[htb]
    \centering
    \caption{\label{tab:PINNs}The results of PINNs solutions for 2D Navier-Stokes equations. w/o is the original version of PINNs. We compare the L2 relative errors of physical variables and unknown parameters, respectively. We also report the total number of sample points of each method.}
    \resizebox{\linewidth}{!}{%
        \centering
        \begin{tabular}{cccc}
        \toprule
        Method & w/o\cite{Raissi2019PINN} & DeepXDE\cite{lu2021deepxde} & \textbf{w(ours)} \\
        \midrule
        $u$ & 5.4e-04  & 4.1e-04   & \textbf{3.6e-04}  \\
        $v$ & 1.9e-03  & 1.5e-03    & \textbf{1.3e-03}  \\
        $p$ & 6.6e-02 & 5.1e-02  & \textbf{1.6e-02}  \\
        $\lambda_1$ & 5.8e-02   & 5.9e-02    & \textbf{4.3e-02}  \\
        $\lambda_2$ & 5.0e-03 & 4.7e-03  & \textbf{3.9e-03}  \\
        $N$ & 5000 & 5000  & \textbf{3920}    \\
        \bottomrule
        \end{tabular}
    }
\end{table}

As shown in \autoref{tab:PINNs}, our octree-based sampling optimization achieves improvements of the original PINNs by $33.33\%$ on $u$, $31.58\%$ on $v$, $75.76\%$ on $p$, $25.86\%$ on $\lambda_1$, and $22.00\%$ on $\lambda_2$, respectively. At the same time, only $78.40\%$ of sample points are required. It indicates that our method produces higher solution accuracy with fewer sample points. In addition, compared to DeepXDE, our octree-based sampling optimization can provide better importance sampling on training PINNs.

\section{Limitations and Discussions}\label{sec:limitations}
As described in \autoref{sec:RBProblem} and \autoref{sec:ablationstudy}, our octree-based sampling optimization is able to greatly reduce the number of sample points required to achieve a specific level of convergence. However, since the indispensable octree reconstruction relies on the time-consuming computation of the loss distribution, the saving of training time is not as dramatic as sample points. Training time is the most concern for processing massive scientific data. We must explore how to accelerate the estimation of the loss distribution. As mentioned in \autoref{sec:reconstruction}, we currently downsample the training data to reduce the time for rebuilding the loss distribution. But a larger downsampling factor will affect the performance of the reconstructed octree. Besides, the minimum number of epochs for each stage of training is a key hyperparameter of our method. How to determine it reasonably according to the specific application is a problem that needs to be further solved.

\section{Conclusion}\label{sec:conclusion}
We propose a novel octree-based hierarchical sampling optimization for the volumetric super-resolution of scientific data. It effectively implements importance sampling by exploiting the importance distribution of training data. We model the volumetric scientific data by leveraging its importance distribution using octree, wherein each leaf node corresponds to an indivisible subblock in the volumetric domain. In particular, we reconstruct the octree whenever the loss reaches a bottleneck, avoiding failure of the sampling optimization. We conduct adequate experiments to show that our method greatly reduces the number of sample points required to achieve a specific level of convergence. Besides, our method can also be used in training tasks of other deep neural networks like PINNs. Future work will focus on the acceleration of the loss distribution computation. We hope that the octree-based sampling optimization will facilitate research on accelerating the training of deep learning models.

\section*{Acknowledgements}
Xinjie Wang was supported by the National Key R\&D Program of China (\# 2022YFC2803805), the Fundamental Research Funds for the Central Universities (\# 202313035), the Shandong Provincial Natural Science Foundation of China (\# ZR2021QF124), and the China Postdoctoral Science Foundation (\# 2021M703031). Xiaogang Jin was supported by the National Natural Science Foundation of China (\# 62036010) and the Key R\&D Program of Zhejiang Province (\# 2022C03126).

% \section*{References}

\bibliography{main.bib}

\end{document}